\documentclass[prb,showpacs,twocolumn,floatfix]{revtex4}
\usepackage{graphicx}
\begin{document}
\def\rhov{{\mbox{\boldmath{$\rho$}}}}
\def\tauv{{\mbox{\boldmath{$\tau$}}}}
\def\Lambdav{{\mbox{\boldmath{$\Lambda$}}}}
\def\sigmav{{\mbox{\boldmath{$\sigma$}}}}
\def\xiv{{\mbox{\boldmath{$\xi$}}}}
\def\oh{{\scriptsize 1 \over \scriptsize 2}}
\def\of{{\scriptsize 1 \over \scriptsize 4}}
\def\tf{{\scriptsize 3 \over \scriptsize 4}}
\title{Cluster Probability in Bootstrap Percolation}

\author{A. B. Harris and Andrea J. Liu}

\affiliation{ Department of Physics and Astronomy, University
of Pennsylvania, Philadelphia, PA 19104}
\date{\today}

\begin{abstract}
We develop a recursive formula for the probability of a $k$-cluster
in bootstrap percolation.
\end{abstract}
\maketitle

\section{INTRODUCTION}
The model of bootstrap percolation~\cite{CHALUPA} (or $k$-core percolation,
as we will refer to it) has a connection to jamming transitions,
including the glass transition.~\cite{BOOK}   Many features of
dynamical arrest in glassforming liquids are captured by
kinetically-constrained spin models,~\cite{RITORT} which map 
onto $k$-core percolation and its variants.  In addition, Schwarz,
et al.~\cite{SCHWARZ} have argued that there is an analogy (but not
a strict mapping) between $k$-core percolation and the jamming
transition of frictionless granular packings~\cite{OHERN,SILBERT}, 
similar to the analogy between $k$-core percolation and rigidity 
percolation.~\cite{MOUKARZEL}   In the mean-field, or 
infinite-dimensional limit, the $k$-core percolation transition
was shown to be peculiar in Ref. \onlinecite{SCHWARZ} and
subsequently in Ref. \onlinecite{MENDES}.  The transition is
discontinuous,~\cite{CHALUPA,PITTEL} but also exhibits power-law
scaling~\cite{CHALUPA,PITTEL} and a diverging susceptibility and
length scale~\cite{SCHWARZ}.  This unusual type of transition
has been called a random first-order phase transition.~\cite{KIRKPATRICK}
What makes the mean-field $k$-core percolation transition 
particularly interesting is that its exponents are identical to the 
mean-field exponents of several models that have been proposed for 
jamming transitions,~\cite{KIRKPATRICK,CHAKRABORTY,BIROLIBOUCHAUD,TBF}
raising the possibility that $k$-core percolation may be in the
same universality class as these models.

For jamming transitions, the random first-order nature of the transition
does not appear to be an artifact of the mean-field approximation.
Numerical simulations~\cite{OHERN,SILBERT} and scaling 
arguments~\cite{WYART1,WYART2} suggest that the jamming transition of
frictionless granular packings (Point J) is discontinuous with a diverging
length scale in two and three dimensions.  It has also been argued
that the finite-dimensional glass transition, if it exists as a true
phase transition, has a mixed character, with a discontinuity in the
infinite-time limit of the dynamical structure factor and a diverging
dynamical length scale.~\cite{KTW,TONINELLI}  This raises the question
of the nature of the $k$-core percolation transition in finite
dimensions $d$.  Expansions in $1/d$ indicate~\cite{ABHJS} that
that the mixed nature of the transition should persist for a range of
large but finite $d$.  In two dimensions, specific variants of
$k$-core percolation have been identified that are rigorously known to
exhibit discontinuous transitions with diverging length scales, both
at $p=1$ (full occupancy)~\cite{DAWSON,TBF} and $p<1$ (partial 
occupancy).\cite{TONINELLI}  Numerical studies of other 
variants~\cite{SCHWARZ} also point to the
possibility of mixed transitions in finite dimensions at $p<1$.   
However, it is also known that some realizations of the model,
such as $k=3$ on the two-dimensional triangular lattice, 
exhibit continuous transitions.~\cite{CHAVES}  Thus, in finite 
dimensions, it appears that different versions of
$k$-core models can exhibit at least three different types of
transitions:  continuous transitions, mixed transitions at $p=1$
and mixed transitions at $p<1$.  

In order to sort through this confusing state of affairs, it would
be useful to map $k$-core percolation onto a spin model, which could
then be studied by a wide variety of methods.  Such an approach
proved extremely powerful for ordinary percolation.  A necessary 
first step towards this goal is to be able to count $k$-core clusters
with the correct probabilities.  In this paper, we 
show how this can be done by deriving
a relation for the probability distribution of clusters.
 
The $k$-core ensemble is defined as follows.
We start with the ensemble of $2^{\cal N}$ percolation configurations
on a {\it finite} lattice of ${\cal N}$ bonds.  Each percolation configuration
${\cal C}_p$ has probability 
\begin{eqnarray}
P_p({\cal C}_p)  &=& p^{n_{\rm occ}({\cal C}_p)} (1-p)^{n_{\rm vac}({\cal C}_p)}
\ ,
\end{eqnarray}
where $n_{\rm occ}({\cal C}_p)$ and $n_{\rm vac}({\cal C}_p)$ are, 
respectively, the number of occupied and vacant bonds in the percolation
configuration ${\cal C}_p$. A $k$-core configuration ${\cal C}_k$
is obtained from ${\cal C}_p$ by removing all bonds
which intersect a site to which there are less than $k$ remaining 
bonds. This process is recursively continued because as bonds are
so removed, they may cause more sites to have less than $k$ neighbors.
After this culling, the remaining $k$-cluster is one in which all sites
have at least $k$ occupied bonds.  The weight of the configuration
${\cal C}_k$ is the sum of the percolation probabilities over all
the percolation configurations which gave rise to ${\cal C}_k$.

In the percolation problem one can easily establish that the probability
of finding an isolated cluster\cite{GAMMA} $\Gamma_p$ is given by
\begin{eqnarray}
P_p(\Gamma_p) &=& p^{n(\Gamma_p)} (1-p)^{t(\Gamma_p)} \ ,
\label{EQ2} \end{eqnarray}
where $n(\Gamma_p)$ is the number of occupied bonds in the cluster
$\Gamma_p$ and $t(\Gamma_p)$ is the number of perimeter bonds which
must be unoccupied in order to define the limit of the percolation
cluster $\Gamma_p$.  One would like to write a similar formula for
$P_k(\Gamma)$, the probability of
a $k$-cluster $\Gamma_k$, but it is obvious that this is not quite so
trivial because of the complications of the culling process.  It is
the purpose of this paper to develop such a formula.

\section{FORMULATION}

To proceed we introduce the function $e_\Gamma({\cal C}_p)$ which is
defined as unity if the percolation configuration ${\cal C}_p$,
when culled, gives rise to the rooted $k$-cluster $\Gamma$.  A rooted
cluster includes the site at the origin and two rooted clusters
are distinct unless their list of occupied bonds coincides exactly.
The final culled configuration must contain the rooted $k$-cluster
$\Gamma$, but can also contain an arbitrary number of $k$-clusters
disconnected from $\Gamma$.  If culling of ${\cal C}_p$ does not produce
a configuration containing the rooted cluster $\Gamma$, then
$e_\Gamma ({\cal C}_p)$ is zero.  With this definition is seems obvious
that
\begin{eqnarray}
P_k(\Gamma) &=& \sum_{{\cal C}_p} e_\Gamma ({\cal C}_p) P_p({\cal C}_p) \ .
\end{eqnarray}
This relation expresses the fact that the cluster $\Gamma$ inherits the
total percolation probability of all configurations which contain the
cluster $\Gamma$ after culling. Note that a fixed $\Gamma$ inherits
probability from not only single clusters which, when culled, yield
$\Gamma$, but also from the vast array of configurations which, when culled,
yield $\Gamma$ together with arbitrary disconnected $k$-clusters. 
Now it is clear that the sum over ${\cal C}_p$ can be restricted to
configurations which either are equal to $\Gamma$ or include it as
a proper subset. So we write
\begin{eqnarray}
P_k(\Gamma) &=& \sum_{{\cal C}_p : {\cal C}_p \geq \Gamma}
e_\Gamma ({\cal C}_p) P_p({\cal C}_p) \nonumber \\ &=&
\sum_{{\cal C}_p : {\cal C}_p \geq \Gamma} 
\biggl(1 - [1- e_\Gamma ({\cal C}_p)] \biggr)  P_p({\cal C}_p) \ .
\end{eqnarray}
But
\begin{eqnarray}
\sum_{{\cal C}_p : {\cal C}_p \geq \Gamma} P_p({\cal C}_p ) &=& p^{n(\Gamma)} 
\end{eqnarray}
because after we occupy the bonds of $\Gamma$ (with probability 
$p^{n(\Gamma)}$) we sum over all states of all the bonds not
in $\Gamma$ (which are occupied with probability $p$
and vacant with probability $1-p$). Thus
\begin{eqnarray}
P_k(\Gamma) &=& p^{n(\Gamma)} -
\sum_{{\cal C}_p : {\cal C}_p \geq \Gamma} 
[1- e_\Gamma ({\cal C}_p)] P_p({\cal C}_p) \ .
\end{eqnarray}
The result of Eq. (\ref{EQ2})
for the probability of a percolation cluster can be
calculated in this same style. Note the meaning of the factor
$[1-e_\Gamma ({\cal C}_p)]$: it is an indicator function such
that culling does
{\it not} lead to the $k$-cluster $\Gamma$.  But since ${\cal C}_p$
contains $\Gamma$ but does not cull to $\Gamma$, it must be that the
configuration ${\cal C}_p$ leads to a cluster $\Gamma_>$ which
includes $\Gamma$ as a proper subset.  So we have
\begin{eqnarray}
P_k(\Gamma) &=& p^{n(\Gamma)} -
\sum_{\Gamma': \Gamma' > \Gamma} \sum_{{\cal C}_p : {\cal C}_p \geq \Gamma'} 
e_{\Gamma'} ({\cal C}_p) P_p({\cal C}_p) \nonumber \\ &=&
p^{n(\Gamma)} - \sum_{\Gamma' : \Gamma' > \Gamma} P_k(\Gamma') \ .
\label{RES} \end{eqnarray}
Note that the sum over $k$-clusters $\Gamma'$ does {\it not} include
$\Gamma'=\Gamma$.  Equation (\ref{RES}) is the principal result of this
paper.  On a qualitative level we can see the following: for
percolation Eq. (\ref{EQ2}) gives what could be called
a perimeter renormalization $(1-p)^{n_t(C)}$, whereas the subtraction
terms in Eq. (\ref{RES}) give a much
weaker ``culling renormalization.''  In the extreme case of lattice
animals, there is no perimeter renormalization at all. In that case
$P(C)=p^{n(C)}$ (where $p$ is interpreted to be the bond fugacity,
usually denoted $K$.).  The continuous aspect of $k$-core percolation
has\cite{SCHWARZ} the same anomalous correlation length exponent $\nu=1/4$
as does lattice animals in infinite dimensions.\cite{LUB}  A heuristic explanation of this is
that the culling renormalization is probably closer to the absence
of a perimeter renormalization in lattice
animals than to the perimeter renormalization of percolation.

Now we briefly explore some consequences of the above result.  
We iterate Eq. (\ref{RES}) to get
\begin{eqnarray}
P_k(\Gamma_k) &=& p^{n(\Gamma_k)}
- \sum_{\Gamma_k^{(1)} > \Gamma_k} p^{n(\Gamma_k^{(1)})} 
+ \sum_{\Gamma_k^{(2)} >\Gamma_k^{(1)} > \Gamma_k} p^{n(\Gamma_k^{(2)})}
\nonumber \\
&& \dots + (-1)^m \sum_{\Gamma_k^{(m)} > \Gamma_k^{(m-1)} \dots >
\Gamma_k^{(1)} > \Gamma_k} p^{n(\Gamma_k^{(m)})} + \dots \nonumber \\
&=& p^{n(\Gamma_k)} + \sum_{m=1}^{m_{\rm max}} (-1)^m \sum_{\{\Gamma_k{(m)}\}}
p^{n(\Gamma_k^{(m)})} \ ,
\end{eqnarray}
where, for simplicity, we do not write out (in the last line)
the inclusions.  Because we are dealing with a finite lattice, the
index $m$ is bounded by a maximum value, denoted $m_{\rm max}$.

Next we consider the calculation of the connectedness
susceptibility, $\chi$, which can be written as
\begin{eqnarray}\
\chi &=& \sum_{\Gamma_k} P_k(\Gamma_k) n(\Gamma_k)^2 \ .
\end{eqnarray}
Thus we write
\begin{eqnarray}
\chi&=& \sum_{\Gamma_k} n(\Gamma_k)^2 \biggl[
p^{n(\Gamma_k)} \nonumber \\
&& \ \ \ \ \ + \sum_{m=1}^{m_{\rm max}} (-1)^m \sum_{\{\Gamma_k{(m)}\}}
p^{n(\Gamma_k^{(m)})} \biggr] \ .
\end{eqnarray}
By relabeling clusters, this can be rearranged to combine all terms
with the same factor $p^{n(\Gamma )}$, in which case one has
\begin{eqnarray}
\chi&=& \sum_{\Gamma_k} p^{n(\Gamma_k)} \biggl[
n(\Gamma)^2  - \sum_{\Gamma_k^{(1)}< \Gamma} [n(\Gamma_k^{(1)})]^2 
\nonumber \\ && \
\dots + (-1)^m \sum_{\Gamma_k^{(m)} < \Gamma_k^{(m-1)} \dots <
\Gamma_k^{(1)} < \Gamma} [n(\Gamma_k^{(m})]^2
\nonumber \\ && \  + \dots \biggr]  \ .
\label{chi}
\end{eqnarray}
This result no longer requires that ${\cal N}$ be finite
because for any finite cluster $\Gamma_k$, the number of terms coming
from included subclusters in the sum inside the square brackets is
independent of system size.  Thus this formula presumably can be
used as long as $p< p_k$
and it may yield a possible route to an analytic approach to
the calculation of $\chi$ for finite dimensional lattices.

The result Eq.~ (\ref{chi}) can be generalized to any observable
$G(\Gamma_k)$ that can be defined for a k-cluster $\Gamma_k$.
In that case, the ensemble-averaged value of $G$ is
\begin{equation}
\langle G \rangle = \sum_{\Gamma_k}p^{n(\Gamma_k)} G_c(\Gamma_k),
\label{Gave} 
\end{equation}
where the cumulant function $G_c$ is
\begin{eqnarray}
G_c(\Gamma_k)&=&G(\Gamma_k)-\sum_{\Gamma_k^{(1)}< \Gamma_k}G(\Gamma_k^{(1)}) 
\nonumber \\ && \
\dots + (-1)^m \sum_{\Gamma_k^{(m)} < \Gamma_k^{(m-1)} \dots <
\Gamma_k^{(1)} < \Gamma_k} G(\Gamma_k^{(m)})
\nonumber \\ && \  + \dots
\label{Gcdef}
\end{eqnarray}
or more simply,
\begin{equation}
G_c(\Gamma_k)=G(\Gamma_k)-\sum_{\Gamma_k^{(1)}< \Gamma_k}G_c(\Gamma_k^{(1)}) .
\end{equation}
This cumulant subtraction for the $k$-core percolation problem is much 
more difficult to handle than it is for ordinary percolation.

We thank Vincenzo Vitelli for stimulating discussions and a critical
reading of the manuscript.  This work was supported by NSF-DMR-0605044 
(AJL).


\begin{thebibliography} {99}
\bibitem{GAMMA}
Single clusters are denoted by $\Gamma$ with appropriate scripts.
Configurations (which may consist of a union of clusters) are
denoted ${\cal C}$ with appropriate scripts.
\end{thebibliography}
\end{document}